\documentstyle[prb,aps,epsf]{revtex}

\begin{document}
\draft

\twocolumn[\hsize\textwidth\columnwidth\hsize\csname @twocolumnfalse\endcsname


\title
{\bf
Amorphous indium phosphide from first principles
}

\author{Laurent J.\ Lewis,\cite{byline1} Alessandro De Vita, and Roberto Car}

\address{
Institut Romand de Recherche Num\'erique en Physique des Mat\'eriaux (IRRMA),
IN-Ecublens, CH-1015 Lausanne
}

\maketitle

\begin{center}
{\bf Submitted to Physical Review B} \\ \today
\end{center}

\begin{abstract}

We report detailed and extensive first-principles molecular-dynamics (MD)
simulations of the structure and electronic properties of amorphous InP
produced by rapid quenching from the liquid. The structure of the material is
found to be strongly ordered chemically, even though there is a significant
number of coordination defects and despite the presence of odd-membered
rings. We find, as a consequence, that there exists ``wrong bonds'' in the
system, in an amount of about 8\%; these result from the presence of
coordination defects, {\em not} of local composition fluctuations, as has
been conjectured. The system, in fact, is found to be over-coordinated, which
might be the reason for the observed higher density of {\em a-}InP compared
to {\em c-}InP. We have also investigated the possibility of
pressure-amorphizing InP. Our calculations indicate that the cost of a
transformation of the compressed zinc-blende crystal into an amorphous phase
is so large that it is very unlikely that it would take place.

\end{abstract}

\pacs{PACS numbers: 61.43.Bn, 61.43.Dq, 64.70.Pf, 71.23.Cq}

\vskip2pc]

\narrowtext

\section{Introduction}\label{intro}

Despite considerable work over the last few decades, precise
understanding of the static and dynamic structure of glasses and amorphous
materials remains a challenge to theorists and experimentalists.
\cite{zachariasen32,lannin87,elliott83,elliott89,yonezawa96} Only average
properties of these materials are accessible to experiment; even in such a
simple material as {\em a-}Si, a covalent semiconductor, detailed
experimental knowledge of the atomic arrangements on the local lengthscale is
missing. The average coordination number of {\em a-}Si, for instance, is not
known exactly, though it appears that it is almost the same as for {\em
c-}Si, i.e., 4.\cite{khalid} The only way of obtaining detailed microscopic
information on the local atomic structure is thus via theoretical modelling.
In particular, {\em ab-initio} molecular-dynamics (MD) simulations, which
describe accurately the interatomic potentials, have been able to generate
structural models of {\em a-}Si and {\em a-}GaAs which yield measurable
quantities in good agreement with experiment.\cite{stich,fois} {\em
Ab-initio} calculations are, however, computationally very demanding.
Empirical potentials such as Stillinger-Weber\cite{stillinger85} or
Tersoff\cite{tersoff89} work reasonably well for Si, Ge and their alloys, but
there exist no such models for III-V compounds. Indeed, these materials are
inherently much more difficult to model than the corresponding elemental
systems because of the added complexity of (partly) ionic bonding, which
results in a strong degree of chemical order in the crystal. Despite
these difficulties, a set of transferable tight-binding (TB) potentials has
been developed for some of the III-V's, in particular GaAs and
GaP.\cite{molteni94a} In recent MD studies of {\em
a-}GaAs\cite{molteni94b,seong96,mousseau97} and {\em l-}GaAs,\cite{molteni93}
these models were found to produce results in good agreement with those from
{\em ab-initio} studies and from experiment.

InP is an important material for the industry of microelectronic and
optoelectronic devices, in particular in the field of high-speed computing
and communications.\cite{coutts85} In spite of its potential, it has been
much less studied than GaAs (the two materials have similar bandgaps), mostly
because of the difficulty in fabricating high-quality InP crystals in large
enough quantities. Resorting to the amorphous phase of the material might be
a way out of this problem; indeed, {\em a-}InP is expected to find its way in
the fabrication of integrated circuits.

Attempts at fitting a reasonable tight-binding model for InP have so far been
unsuccessful and, as mentioned above, there exists no empirical or other
potential for this material. A first-principles approach therefore seems to
be the only possible avenue for constructing models of the amorphous
material. Here we propose a model for stoichiometric {\em a-}InP obtained by
a melt-and-quench cycle. To our knowledge, this constitutes the very first
attempt at constructing a realistic model of {\em a-}InP. Experimentally, the
structure of the material remains to a large extent unresolved, despite
the fact that some structural measurements have been reported in the
literature (see below). Several questions pertaining to the local atomic
order remain open. In particular, though it is clear that {\em a-}InP is
disordered from both chemical and structural viewpoints, experiment has not
yet given a precise value for the proportion of wrong bonds in the material
--- and its relation to coordination fluctuations. Thus, for instance, it is
not clear if wrong bonds result from the presence of topological defects,
such as odd-membered rings, or from local compositional fluctuations (i.e.,
antisites or clustering) arising from conditions of
preparation.\cite{gheorghiu83} In the case of InP, which is rather strongly
ionic, heteropolar bonding should be favored over homopolar bonding, and the
proportion of wrong bonds consequently reduced.

{\em a-}InP is normally produced by flash evaporation of {\em c-}InP and
deposition onto an appropriate substrate,
\cite{gheorghiu83,ouchene83,flank87,udron92,theye87} but it can also be
obtained by ion implantation --- which in principle yields better-quality
material with reproducible properties --- though usually not in quantity
sufficient for such atomic structural measurements as x-rays to be carried
out.\cite{roorda96} There are only very few reports of
ion-implantation-amorphization of InP.\cite{wendler94} Other covalent
semiconductors, such as Si and GaAs, can also be produced through proper
pressure treatment.\cite{vohra90,tsuji93,tsuji95,sidorov94} To our knowledge,
{\em a-}InP has never been produced in this way. It has however been
conjectured that pressure-induced amorphization should not occur in
strongly-ionic compound semiconductors.\cite{tsuji95} This conjecture has not
been verified; it is clearly of interest to examine the question in the case
of InP, which is significantly more ionic than GaAs (0.421 vs 0.310 --- cf.\
Ref.\ \onlinecite{phillips73}).

We have therefore also examined, in the course of this study, the possibility
of fabricating {\em a-}InP through the application of pressure. Our computer
simulations indicate that InP does {\em not} amorphize under pressure, even
for values largely in excess of those required for the system to transform
into the high-pressure NaCl phase. The energy of the compressed zinc-blende
crystal, we find, remains lower than that of the amorphous phase produced
from the melt, {\em even at higher density.} Thus, there is no chance for
amorphization to take place: the cost of breaking the strong ionic bonds is
just too large. In fact, the system is found to undergo a transition to a
complex --- but ordered --- phase that maintains the chemical order of the
system, i.e., that introduces no wrong bonds. Thus, it appears that, indeed,
strongly-ionic materials are {\em not} good candidates to pressure-induced
amorphization: more ``violent'' processes are required, and this suggests
that the amorphous phase cannot exist in absence of wrong bonds.

There is definite experimental evidence that MeV-ion-bombarded {\em c-}InP
contracts with respect to equilibrium material.\cite{cliche94} The density of
{\em a-}InP, in fact, is slightly (a fraction of a \%) {\em larger} than that
of {\em c-}InP, which is a bit surprising in view of the fact that both {\em
a-}Si and {\em a-}GaAs are less dense than their crystalline
counterparts.\cite{laaziri94,ascheron92} Our calculations are consistent with
this observation in that the system is found to be, on average,
overcoordinated, while the average bond length is larger than in the crystal.
In contrast, using {\em ab-initio} MD and TB-MD, we have found {\em a-}GaAs
--- which is less dense than {\em c-}GaAs --- to be undercoordinated, albeit
only slightly.\cite{fois,seong96,mousseau97}

Our paper is organized as follows. In Sec.\ \ref{back}, in order to put our
contribution in proper perspective, we present a summary of the information
known from both experiment and theory on {\em a-}InP. In Sec.\ \ref{comp}, we
provide details of our computational framework, including a description of
the ground state (zinc-blende) and high-pressure (NaCl) phases. Discussion of
our results for the model prepared by melt-and-quench is given in Sec.\
\ref{mandq}. There, we first present the results for the liquid phase, in
particular static structure and diffusion. The structure and properties of
the amorphous, low-temperature, phase is discussed next in terms of radial
distribution functions, static structure factors, distribution of bond and
dihedral angles, coordination numbers and bonding characteristics,
vibrational properties, and density of electron states. In Sec.\
\ref{pressure}, finally, we present our findings on the possibility of
amorphizing InP through the application of pressure.

\section{Background}\label{back}

\subsection{Experiment}\label{backE}

To our knowledge, only very few X-ray or EXAFS,
\cite{ouchene83,flank87,udron92} and only one electron
diffraction,\cite{gheorghiu83} experiments of {\em a-}InP (prepared by flash
evaporation and deposition, and usually non-stoichiometric) have been
reported; the accuracy of these meaurements, as we discuss now, does not
allow a precise determination of the local atomic order. The measured
nearest-neighbour distances and coordination numbers for the various types of
correlations are listed in Table \ref{exp_st}. The error bar on the
nearest-neighbour distances is reported to be $\pm$0.02--0.05 \AA, while on
the coordination numbers, these are of the order of $\pm$0.4--0.5, but this
also depends on the model used to fit the EXAFS data, as can be seen in Table
\ref{exp_st} for In$_{33}$P$_{67}$.

The proportion of wrong bonds in the above measurements is reported to be
anything between 10 and 40 \% (after taking care of the off-stoichiometry of
the samples). The origin of the wrong bonds is not at all clear; part of the
problem arises because of variations of composition (which, at this level,
can be {\em very} significant): while Flank {\em et al.}\cite{flank87}
believe that the system partly phase separates (i.e., clustering of the
excess P takes place), Udron {\em et al.}\cite{udron92} indicate that the P
is more-or-less uniformly distributed in the samples. It is also suggested
that wrong bonds are due to local composition fluctuations rather than the
presence of topological defects,\cite{gheorghiu83,ouchene83} and in
particular odd-membered rings.

The total coordination number of each species is found in general to be quite
close to 4, as can be seen in Table \ref{exp_st} (summing the partial
coordinations), and this is also consistent with core-level-shift
measurements.\cite{ouchene83} It is therefore tempting to conclude that
odd-membered rings are absent in {\em a-}InP (though, of course, a perfect
{\em average} total coordination of 4 does not preclude the existence of
odd-membered rings) and, likewise, that wrong bonds are absent. However,
using the same set of EXAFS data but different fitting schemes, Flank {\em et
al.}\cite{flank87} found the coordination of In to be either 4 or 4.8, as
indicated in Table \ref{exp_st}. The error bar on these numbers is {\em very}
significant and evidently prevents firm conclusions from being drawn. In
fact, based on the ``measured'' coordination numbers, Flank {\em et al.} find
a proportion of wrong-bonded In atoms of almost 40\%, but only 10\% for P; taking
the stoichiometry of the sample into account, they indicate that their
fitting model must be incorrect. One must conclude, therefore, that the error
bar on the experimental values of the partial coordination numbers is so
large that, for all practical purposes, they are at present unknown.

Electron diffraction experiments have also been performed\cite{gheorghiu83}
on {\em a-}InP samples also prepared by flash evaporation and deposition.
Though such quantities as partial coordination numbers are not provided, a
detailed analysis of the total pair correlation function of the material
suggests that {\em a-}InP is ``more disordered'' than {\em a-}Ge and other
III-V's: the first nearest-neighbour distance is larger that its crystalline
counterpart, with a rather wide spread in the distances; the second peak is
shifted towards smaller values, indicating that the average bond angle is
smaller; finally, the third peak is more-or-less buried in the background,
suggesting that order is totally lost beyond second nearest-neighbours.

These electron diffraction measurements were interpreted in terms of
``standard'' ball-and-stick, continuous-random-network models, namely those
of Polk\cite{polk71} and Connell and Temkin.\cite{connell74} The Polk model
contains odd-membered rings, while the Connell-Temkin model does not. The
measured structure of {\em a-}InP seems to be more adequately described by
the {\em unrelaxed} Connell-Temkin model, i.e., without odd-membered rings,
in line with the above remarks. (This also agrees with a recent study of the
structure of {\em a-}GaAs, as discussed in the next section). It is expected,
however, that relaxation of the Connell-Temkin model would bring about
odd-membered rings.

\subsection{Theory/Models}\label{backT}

To our knowledge, no structural model specific to {\em a-}InP has been ever
been proposed. Only generic ball-and-stick models (Polk, Connell-Temkin) have
been used to interpret strutural data; no computer model, based on any kind
of potential, has been reported.

Based on such a generic model, the density of electron states has been
calculated by O'Reilly and Robertson.\cite{oreil86} It has been found that
wrong bonds, presumably the most important type of defects in this material,
lead to a significant number of states in the gap, and are therefore
extremely important in determining the electronic properties of the material.
We will bring additional evidence for this in the present paper.

It is appropriate to mention at this point that an optimized model for
another III-V compound, {\em a-}GaAs, was very recently developed by one of
the authors (LJL) and a collaborator.\cite{mousseau97} Using the new
``activation-relaxation technique'' (ART)\cite{barkema96} for relaxing
complex structures at 0 K, a model was built that possesses almost perfect
coordination and is essentially free of wrong bonds. In this study, it was
demonstrated that a Connell-Temkin-like model, which contains no odd-membered
rings, provides a better description of {\em a-}GaAs than a Polk-type model,
which is more appropriate to elemental semiconductors. Thus, odd-membered
rings must be present in {\em a-}Si and relatively rare in {\em a-}GaAs. How
these conclusions apply to InP, however, is not clear: In addition to
ionicity, InP differs from GaAs in that the atoms are significantly different
in size --- Ga and As belong to the same row of the Periodic Table, while In
and P are two rows apart. Thus, the competition between elastic deformation
energy and Coulomb repulsion will be rather different in the two materials.
It is not possible, at present, to carry out ART simulations in order to
address this issue since there exists no model potentials for InP.

\section{Computational framework}\label{comp}

As mentioned earlier, our calculations were carried out using now standard
first-principles molecular-dynamics~\cite{car85} plane-wave/pseudopotential
methodology in the local-density approximation (LDA),\cite{dft} with the
exchange-correlation term expressed in the Ceperley-Alder
form.\cite{ceperley80} The version of the code we use, however, is an
implementation of it optimized to run on a block of 32 nodes on a
massively-parallel Cray T3D computer located at \'EPFL. As discussed below,
this has allowed us to carry out extremely long runs in comparison to what
would have been possible on a scalar machine.

All calculations were performed on a constant-volume 64-atom supercell for
stoichiometric InP, i.e., 32 In and 32 P atoms. The supercell volume was
however changed ``by hand'' when appropriate (see below). The plane waves
were cutoff in energy at 12 Ry, which proves to be essentially converged as
far as structural properties are concerned, according to our tests (see also
Ref.\ \onlinecite{alves91}). The interaction between electrons and ion cores
is described in terms of norm-conserving, fully-separable {\em ab initio}
pseudopotentials of the Kleinman-Bylander form.\cite{kleinman82} Only the
$\Gamma$ point was used for integrating the Brillouin zone. The program uses
preconditioning,\cite{tassone94} so a rather large timestep of 10.0 a.u.
(about 0.25 fs), with a cutoff ``mass'' of 3.0 a.u., could be used. The
fictitious mass of the electrons was set to 300 a.u. A Nos\'e thermostat, with
a ``mass'' of $4.32\times10^{10}$ a.u.\ was used to control the temperature;
we have verified that the structural and dynamical properties of our systems
are not significantly influenced by this choice.

\begin{figure}
\epsfxsize=7cm
\epsfbox{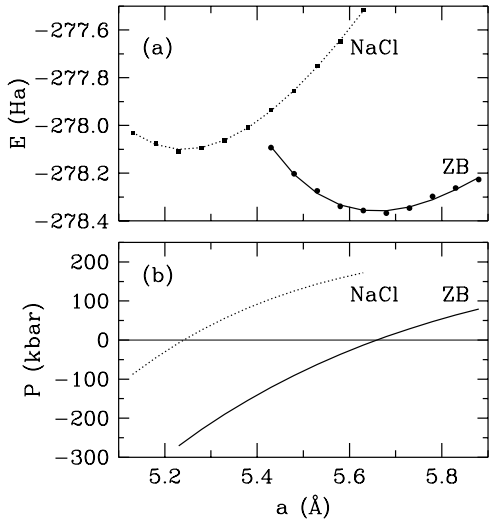}
\vspace{1cm}
\caption{
(a) Total energy and (b) pressure vs lattice parameter for the two structures
considered: ZB and NaCl. The lines are obtained by fitting to Eq.\ (1)
\label{E_P_vs_a}
}
\end{figure}

We have calculated the total energy as a function of lattice parameter using
the above model for both the zinc-blende (ZB, F43m) and the sodium-chloride
(NaCl, Fm3m) structures. The latter structure corresponds to the
high-pressure phase of InP and other III-V compounds.
\cite{minomura62,jamieson63} The results are shown in Fig.\
\ref{E_P_vs_a}(a). The total-energy data were fitted to the ``universal
binding-energy function'' (see, e.g., Ref.\ \onlinecite{wang89})
   \begin{equation}
      E(r) = \alpha (1 + \frac{r-a_0}{\beta}) \exp(-\frac{r - a_0}{\beta})
           + \mbox{constant},          \label{toe_fit}
   \end{equation}
where $a_0$ is the fitted equilibrium lattice parameter and $\alpha$ and
$\beta$ are other fitting parameters. Fig.\ \ref{E_P_vs_a}(b) shows the
pressure ($P=-dE/dV$ at 0 K) for the two phases.

From the data of Fig.\ \ref{E_P_vs_a}(a), we find the lattice parameter to be
5.68 and 5.24 \AA\ for the ZB and NaCl structures, respectively. For the ZB
structure, the calculated lattice parameter is about 3\% smaller than the
experimental value (5.859 \AA). This discrepancy is largely due to our use of
the LDA (which systematically underestimates lattice parameters), and to some
extent also to limited Brillouin-zone sampling: the ``true'' LDA value,
obtained by detailed integration, is 5.74 \AA,\cite{adc} 2\% smaller than
experiment. For the NaCl structure, the computed value for the lattice
constant using the same computational parameters as above is about 9\%
smaller than experiment --- 5.24 vs 5.71 \AA. Accurate bulk calculations on
this phase show that the error originates in part (about 4\%) from
insufficient Brillouin zone sampling ($\Gamma$-point only) and no
Fermi-energy smearing scheme (the NaCl phase is found to be metallic at the
theoretical equilibrium volume). A further 4\% of the error is recovered by
using the non-linear core correction for the exchange-correlation
potential,\cite{froyen} leaving a residual error of about 1\% due to the LDA
and pseudopotential approximations. In the light of these results, we cannot
expect our model to provide an accurate description of this phase with the
run-time calculation parameters reported above, which were required for the
very long production simulations needed (see below). However, since we are
primarily interested in the ZB phase, this will be of relatively little
consequence, and we will still be able to draw qualitative conclusions on the
possibility of pressure-amorphizing InP (Sec.\ V). The energy difference
between the two phases in our calculation is found to be about 0.11 eV/atom
(in favor of ZB), compared to the fully-converged value of 0.15 eV/atom and
to about 0.38 eV/atom from experiment. While the error bar on the
experimental value is not known, it is likely that part of this difference is
due to the LDA approximation.

\section{Melt-and-quench amorphization}\label{mandq}

\subsection{Thermal cycle}\label{therm}

\begin{figure}
\vspace{1cm}
\epsfxsize=7cm
\epsfbox{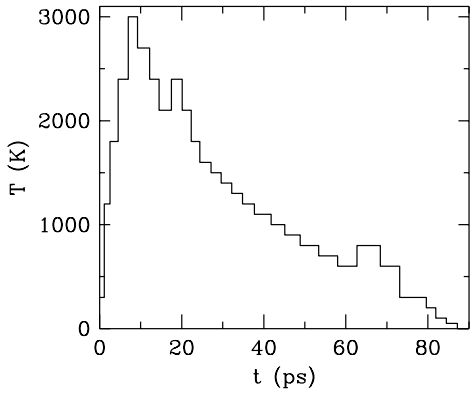}
\vspace{-1cm}
\caption{
Thermal cycle of the melt-and-quench process used to prepare the amorphous
phase, as discussed in the text.
\label{T_hist}
}
\end{figure}

The thermal cycle used to prepare the amorphous sample by melt-and-quench is
summarized in Fig.\ \ref{T_hist}: Starting with a perfect crystal, the system
was first equilibrated at room temperature (300 K), then heated up in steps
until it melted, and finally cooled as slowly as possible into a glass. It
should be stressed that the cooling rate used here, about $2\times 10^{13}$
K/s, is probably the smallest ever achieved in a first-principles simulation
of the liquid-glass transition: the total time covered is a formidable 90 ps,
compared to, typically, $\sim$10 ps in corresponding simulations of other
materials. In spite of this, effects of the finite (and still large) cooling
rate are expected to be present.

For the lattice parameter of the crystalline phase, we used, at all
temperatures, the value obtained above from the 0 K global optimization; it
should be noted that the thermal expansion of {\em c-}InP is very
small,\cite{landolt1,adachi92} and therefore neglecting this effect is of
little consequence. For the liquid, now, the density is larger than that of
the crystal. This quantity is very difficult to calculate in the present
simulation scheme, but is known (approximately) from experiment. Thus we used, for
the liquid, the density of the crystal scaled up by a factor equal to the
experimental ratio of liquid-to-crystalline densities,\cite{landolt2} namely
$\sim$5.1/4.77=1.069. The amorphous phase, finally, is known from experiment
(on ion-implanted {\em a-}InP) to have a density almost exactly equal (to
within 0.5\%) to that of {\em c-}InP.\cite{cliche94} Again, here, this
quantity is very difficult to calculate in the absence of a constant-pressure
option; thus, we simply assumed the amorphous-phase density to be the same as
that of the crystal, an approximation which should be insignificant
compared to other limitations of the study.

Upon heating, the density of the system was changed from that of the crystal
to that of the liquid at 1800 K, i.e., somewhat above the experimental
melting temperature of InP,\cite{landolt1} $T_m = 1335 \pm 50$ K. We found
the system to remain crystalline at this temperature, i.e., to be in a
super-heated state, a consequence of the finite (short) run time. It was then
heated up to 2100 K, and found to melt, and then to 2400 and 3000 K, the
highest temperature considered in this study. After cooling (in steps) to
2100 K, the density was changed back to that of the crystal, and the system
``annealed'' at 2400 K so as to remove the effects of the change in density.
Quenching into the glass was then carried out in steps, proceeding more and
more slowly into structural arrest (see Fig.\ \ref{T_hist}.)

\begin{figure}
\epsfxsize=7cm
\epsfbox{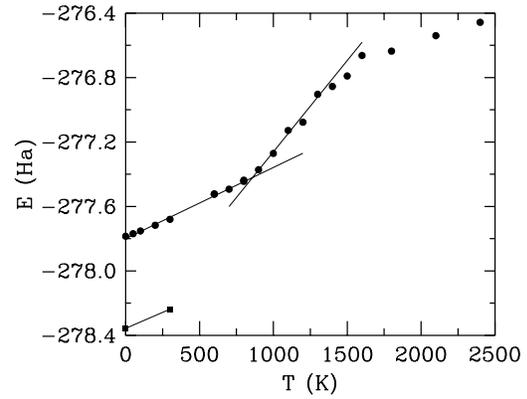}
\vspace{-1cm}
\caption{
Total energy of the system through the liquid-glass transition; here, the
density is that of the crystal. Also shown is the energy of the ZB crystal.
The lines are for guiding the eye.
\label{E_T}
}
\end{figure}

The system was found to remain liquid (non-zero diffusion on the timescale of
the simulations) at temperatures as low as 900 K, indicating a rather strong
hysterisis of the melt-freeze cycle. While this is likely a manifestation of
finite run times, it can also be attributed, in part, to our use of the LDA,
which tends to underestimate the temperature of such transitions: For
instance, in a free-energy calculation of the melting of Si, Sugino and
Car\cite{sugino} found a transition temperature somewhat below (300 K) that
observed experimentally. It is however expected that finite-size effects on
the transition temperature are relatively small. The liquid-glass transition
can be seen very clearly in Fig.\ \ref{E_T}, which shows the total energy
versus temperature upon going through the transition at constant density.

The ground-state energy of the amorphous phase lies approximately 0.24
eV/atom above that of the crystal. This quantity (the latent heat of
crystallization) has to our knowledge never been measured in InP; for Si, it
varies between 0.14 and 0.20 eV/at, depending on the state of relaxation of
the material.\cite{roorda91} Because of the presence of wrong bonds, the heat
of crystallization is expected to be larger in III-V materials than in
elemental semiconductors, consistent with our result.

\subsection{Liquid phase}\label{liquid}

The liquid was studied in detail at four different temperatures: 3000, 2700,
2400, and 2100 K. The calculated diffusion constants are presented in Table
\ref{diff}. The error bar on these numbers is estimated to be of the order of
10\%, arising mainly from the limitations inherent to the method (size and
time). We find no significant differences in the diffusional behaviour of the
two components. From these data, we find an activation energy of about 0.35
eV. To our knowledge, the diffusion constants are not known from experiment
for InP; to give an experimental reference for a comparable
system,\cite{landolt2} in the case of liquid GaAs, $D=1.6\times10^{-4}$
cm$^2$/s at 1550 K, i.e., a bit larger than the values we find here
(extrapolating to lower temperatures).

The structure of the liquid at the various temperatures considered was
analyzed in terms of radial distribution functions, static structure factors,
and coordination numbers. The velocity auto-correlation functions and
distribution of vibrational states were also calculated.

\begin{figure}
\vspace{1cm}
\epsfxsize=7cm
\epsfbox{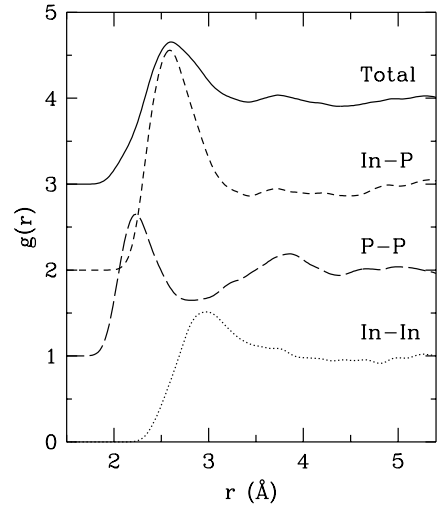}
\caption{
Partial and total radial distribution functions of liquid InP at 2100 K. For
clarity, in this and similar figures, the zeroes are displaced along the $y$
axis.
\label{l_rdf}
}
\end{figure}

The partial radial distribution functions (RDF's) $g_{ij}(r) = \rho_{ij}(r)/
4\pi r^2 c_i \rho_0$ (where $\rho_{ij}(r)$ is the correlation function for
$i$--$j$ pairs, $c_i$ is the relative concentration of species $i$, and
$\rho_0$ is the average number density) provide detailed information about
the short-range arrangements of atoms in the system; they are shown in Fig.\
\ref{l_rdf} for the lowest $T$ examined at the liquid density, viz.\ 2100 K.
We find the liquid, independently of temperature, to have relatively little
structure, essentially restricted to the first or perhaps second
nearest-neighbour peak. Thus, there are essentially no correlations beyond a
distance of about 3.5 \AA, and the ``minimum after the first peak'' is almost
non-existent, except for P-P correlations, which seem to exhibit a
well-defined minimum as well as a second-neighbour peak at this temperature.
As discussed below, this absence of a marked structure will make it rather
difficult to define coordination numbers.

Likewise, we show in in Fig.\ \ref{a_ssf} the partial, $S_{ij}(k)$, and
total, $S(k)$, static structure factors (SSF's) of the liquid at the lowest
temperature. The SSF's are related to the RDF's by a Fourier transform and
are in principle available directly from scattering experiments (neutrons,
x-rays, etc). The SSF's were evaluated directly in reciprocal space in order
to avoid the spurious oscillations that arise in the Fourier transform of a
function that does not terminate smoothly (as is the case for finite-size
models). Just like the radial distribution functions, the static structure
factors show relatively little structure. We know of no experimental {\em
l-}InP data to compare these results with.

The results of Fig.\ \ref{l_rdf} show the most strongly-marked correlation at
short range to consist of In-P hetero-bonding. This is roughly twice as
important as In-In and P-P bonding, which are nevertheless present in very
significant number. Thus, ``wrong bonds'' are very present in this phase (and
of course totally absent in the perfect crystal), very likely a consequence
of the metallic-bonding properties of the liquid, and in qualitative
agreement with the first-principles calculations of {\em l-}GaAs by Zhang
{\em et al}.\cite{zhang90}

\begin{figure}
\vspace{1cm}
\epsfxsize=7cm
\epsfbox{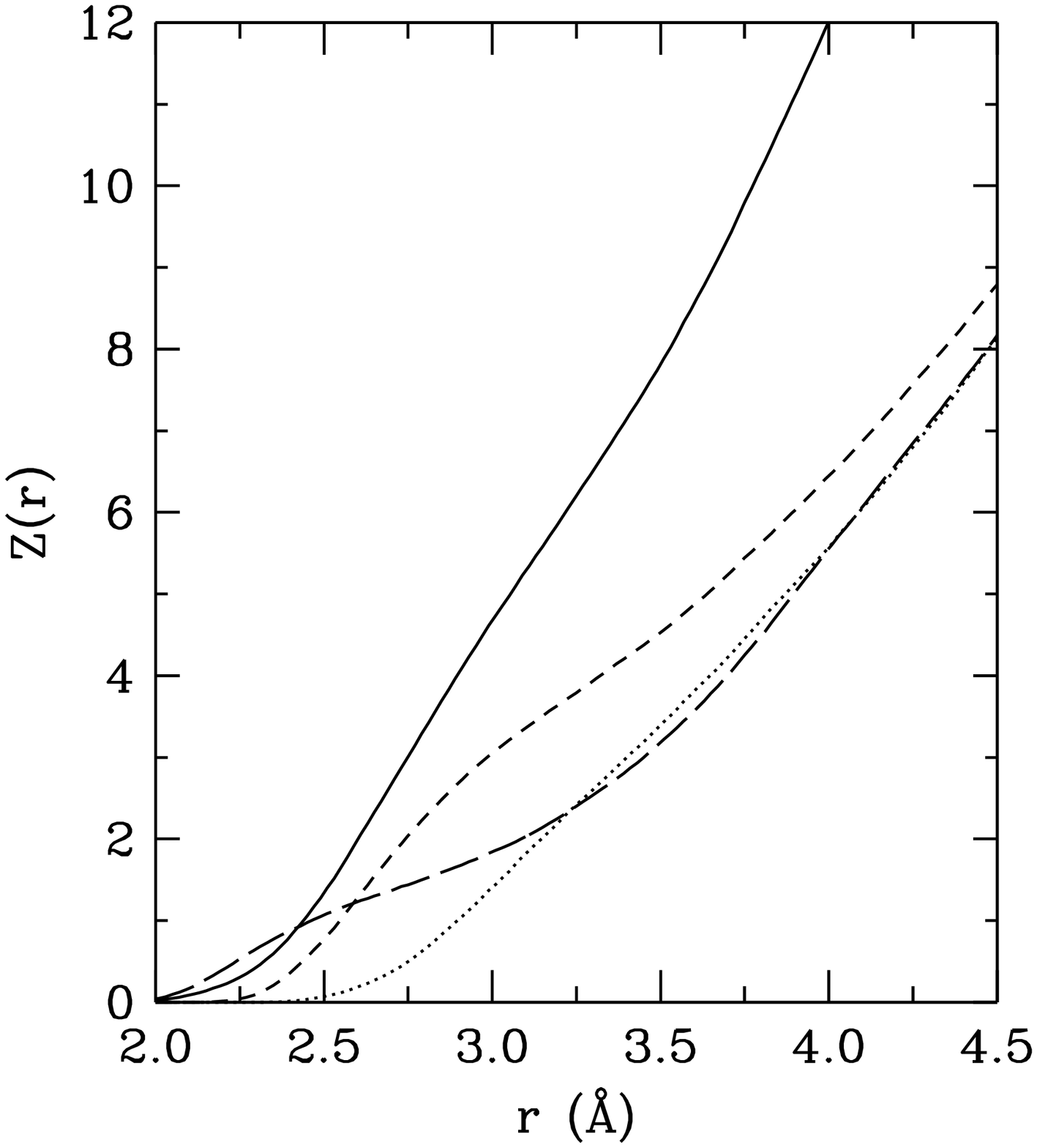}
\caption{
Running coordination numbers corresponding to the radial distribution
functions of Fig.\ \protect\ref{l_rdf}.
\label{l_Zr}
}
\end{figure}

As noted above, defining coordination numbers in such a system is not simple.
We plot in Fig.\ \ref{l_Zr} the ``running coordination numbers'', i.e.,
integrated radial distribution functions, $Z_{ij}(r) = \int_0^r \rho_{ij}(r)
dr$. If coordination numbers were well defined, one would see ``plateaux'' in
these functions, corresponding to the successive neighbour shells, i.e.,
minima in the corresponding radial distribution functions. Clearly there are
no such plateaux here. Nevertheless, we list in Table \ref{tab_Z} the
coordination numbers obtained by choosing some ``reasonable'' first-neighbour
distances (as indicated in the Table).

We find, despite the large error bars, the coordination numbers to
decrease markedly with decreasing temperature, i.e., the covalent character
of the material is increasing upon approaching the transition temperature.
This is true of all three types of partial correlations, and of course also
of the average (total) coordination number. The latter, $Z$, decreases from
8.7 at 3000 K to about 7.0 at 2100 K. We can extrapolate that, at the melting
temperature of InP (1335 K), $Z$ would be about 6.0, as is approximately
found in Si just above melting.

\begin{figure}
\vspace{1cm}
\epsfxsize=7cm
\epsfbox{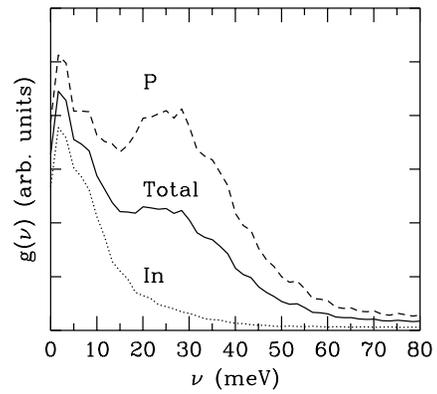}
\caption{
Densities of vibrational states of liquid InP at 2100 K.
\label{l_vdos}
}
\end{figure}

Fig.\ \ref{l_vdos}, finally, gives the density of vibrational states $g(\nu)$
for each atomic species, as well as overall. These were obtained by Fourier
transforming the velocity auto-correlation functions. Though the density of
states for In atoms show essentially no structure --- it decreases rapidly
with frequency --- that for P atoms possesses a shoulder in the 20--40 meV
range. This is likely related to the transverse and longitudinal optical
phonon peaks in {\em c-}InP, respectively at 41 and 45 meV (in the present
model; see below), and is manifest of the fast motion of the light P against
the heavy In atoms. The frequency of this peak should therefore increase upon
decreasing the temperature; indeed, this is is what we find upon examining
$g(\nu)$ at various temperatures (not shown).

\subsection{Amorphous phase}\label{amorph}

\subsubsection{Radial distribution functions and static structure factors}

\begin{figure}
\vspace{1cm}
\epsfxsize=7cm
\epsfbox{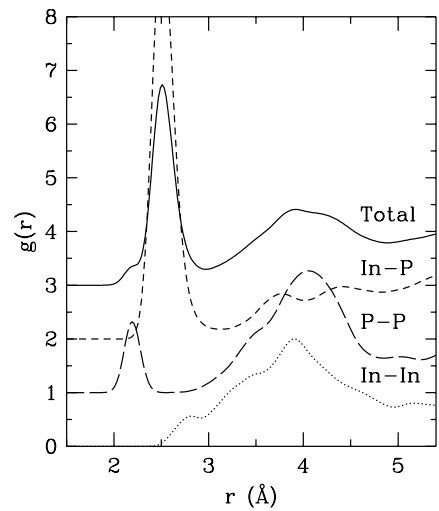}
\caption{
Partial and total radial distribution functions of {\em a-}InP at 300 K.
\label{a_rdf}
}
\end{figure}

The partial RDF's for the fully-relaxed amorphous model at 300 K are
presented in Fig.\ \ref{a_rdf}; also shown is the total (equi-weighted) RDF,
$g(r)$. We observe that the partial In-P RDF is quite similar (in shape) to
the total RDF, reflecting the fact that, as expected, unlike-atom
correlations largely dominate in the amorphous sample at short distances. In
the ideal ZB structure, of course, only hetero bonds are allowed and the
first peak of the total RDF coincides with that of the In-P partial RDF. In
the amorphous material, homo bonds are possible to some extent, even though
hetero bonds prevail, as we discuss below.

The presence of homo bonds is especially evident in the P-P partial
correlation; they manifest themselves as a small peak in the RDF at a
distance of 2.19 \AA, close to the P-P covalent bond distance (2.20 \AA,
twice the covalent radius). This distance is somewhat shorter than the In-P
bond distance (2.51 \AA). In the case of In-In, we observe a shoulder, or
prepeak, at a distance of 2.81 \AA, now larger than the In-P bond distance,
but again close to the covalent bond distance (2.88 \AA). These effects can
clearly be attributed to the size and ionicity differences between the two
species (In is substantially larger than P.) In contrast, in {\em a-}GaAs,
like-atom peaks are found at about the same distance as the unlike-atom peak.
The nearest-neighbour distances we find agree quite closely with those from
experiment reported in Table \ref{exp_st} (though at different chemical
compositions). For P-P, we find 2.19 \AA, vs 2.20-2.24 from experiment; for
In-P, we obtain 2.51 \AA, compared to 2.57--2.59 experimentally; and for
In-In, which is most difficult to define, as is also the case experimentally,
we have 2.81 \AA\ vs 2.76--2.98. We note that part of the observed difference
arises from our model underestimating (by about 3\%) the lattice parameter of
the real material as discussed earlier; in view of this, we conclude that our
model is in close agreement with experiment as far as nearest-neighbour
distances are concerned and {\em modulo} the error bars inherent to both
methods.

In the crystal, the equilibrium LDA In-P bond distance is 2.46 \AA, while
second nearest-neighbours lie at 4.02 \AA. In our amorphous sample, we find,
from the total RDF (Fig.\ \ref{a_rdf}), the nearest-neighbour peak at 2.51
\AA, a bit larger than the corresponding value in the crystal. In contrast,
the second-neighbour peak is at about 3.9 \AA, thus shifted towards smaller
values compared to the crystal, and is much broader. In fact, it is clear
from Fig.\ \ref{a_rdf} that the second peak is made up of at least two
subpeaks, with a shoulder at about 4.4 \AA\ arising from In-P correlations.
In any case, these observations agree with the electron diffraction data of
Ref.\ \onlinecite{gheorghiu83}, discussed in Sec.\ \ref{backE}.

It is clear from Fig.\ \ref{a_rdf} that the concept of nearest-neighbour
distances in the amorphous phase is somewhat ill-defined, especially in the
case of In-In correlations, where the first peak is almost merged into the
second peak. In fact, the ``second-neighbour'' peak, for all correlations, is
rather wide, consisting of several sub-peaks, reflecting the large spectrum
of possible configurations in the disordered phase. All correlations seem to
differ little from unity beyond the second peak, indicating that order, in
the amorphous phase, is indeed very short-range, restricted to the first- and
second-, perhaps third-, neighbour shells. We note, also, that
second-neighbour peaks differ very significantly in shape from the
corresponding peaks in crystalline material.

\begin{figure}
\epsfxsize=7cm
\epsfbox{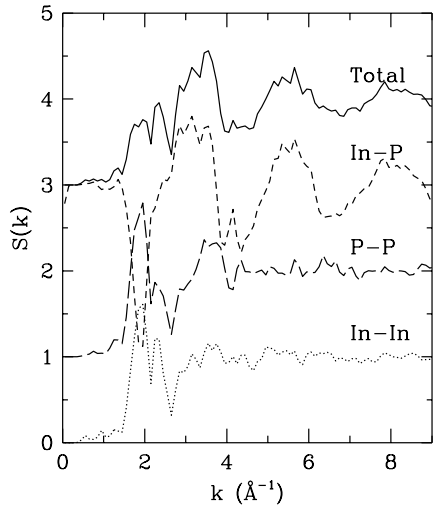}
\caption{
Partial and total structure factors of {\em a-}InP at 300 K.
\label{a_ssf}
}
\end{figure}

For completeness, we present in Fig.\ \ref{a_ssf} the partial and total SSF's
for our model sample. The total SSF was obtained by combining the partial
$S_{ij}(k)$ with equal weights. [In principle, $S(k)$ is a weighted sum of
the partials, where the weights are related to the scattering lengths of the
atoms for the probe used.] The total interference function (essentially the
SSF) of {\em a-}InP at almost stoichiometric concentration, measured by
electron diffraction, has been reported by Gheorghiu {\em et
al}.\cite{gheorghiu83} They observe a small peak at 2.1 \AA$^{-1}$, and three
large peaks at 3.5, 5.7, and 8.0 \AA$^{-1}$, respectively. This correlates
extremely well with the total SSF displayed in Fig.\ \ref{a_ssf}.

\subsubsection{Bond and dihedral angles}

\begin{figure}
\vspace{1cm}
\epsfxsize=7cm
\epsfbox{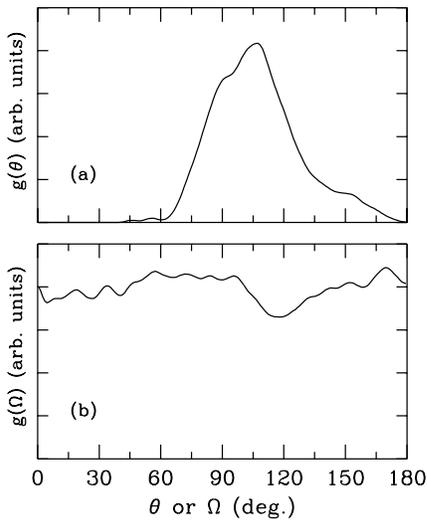}
\vspace{1cm}
\caption{
Distribution of (a) bond and (b) dihedral angles in {\em a-}InP at 300 K.
\label{a_b_dih}
}
\end{figure}

We give in Fig.\ \ref{a_b_dih}(a) the distribution of bond angles in the
amorphous structure, all combinations taken into account. As can be inferred
from the above discussion, the definition of ``bond'' is somewhat arbitrary.
The cutoff distances we used, extracted from the corresponding RDF's (Fig.\
\ref{a_rdf}) are 2.91, 3.13, and 2.55 \AA\ for In-In, In-P and P-P,
respectively; the value for In-In, which hardly exhibits a nearest-neighbour
peak, is subject to a significant error. These cutoff distances will also be
used for determining the coordination numbers, below.

The bond-angle distribution is wide but exhibits a strong peak at about
107$^{\circ}$, a little bit smaller than the tetrahedral angle
(109.5$^{\circ}$). A shift of the bond-angle peak to smaller values has also
been observed by electron diffraction.\cite{gheorghiu83} A similar shift has
been obtained theoretically by Stich {\it et al.} for {\em a-}Si.\cite{stich}
The bond-angle distribution here differs from the case of {\em a-}Si in that
it shows a rather marked shoulder at about 90$^{\circ}$ --- likely arising
from four-membered rings and from those atoms that are five- or six-fold
coordinated --- as well as a weak shoulder at about 150$^{\circ}$, which
perhaps originates from three-fold coordinated atoms. It is quite remarkable
that there exists almost no correlations with an angle of 60$^{\circ}$. This
is in sharp contrast with other tetrahedral semiconductors (elemental or
compound), modeled either empirically or using TB or first-principles MD,
where a significant peak or shoulder is observed at such small angles,
arising from small, e.g., three-membered, rings. This indicates that the
chemistry of this system is robust enough that such defects are rare
(three-membered rings are extremely costly in both elastic-deformation and
electronic-repulsion energies while four-membered rings cost only elastic
energy), as can indeed be verified in Table \ref{ring}, and/or the relaxation
of the present model has been particularly effective. (Ring statistics are
extremely sensitive to details of the local structure, and in particular
coordination; this explains the sizable differences between {\em a-}InP and
{\em a-}GaAs in Table \ref{ring}.)

In Fig.\ \ref{a_b_dih}(b), we give the distribution of dihedral angles
(angles between second-neighbour bonds). In the ZB structure at low
temperatures, the corresponding distribution has two sharp peaks, at 60 and
180$^{\circ}$. In the case of {\em a-}InP, we observe a rather flat
distribution, except for two small dips at 0 and 120$^{\circ}$ (which are
equivalent, on average, for tetrahedral systems), perhaps a memory of the
crystalline phase, but in any case much less pronounced than the
corresponding ones in {\em a-}GaAs,\cite{mousseau97} which chemically orders
a bit more strongly than {\em a-}InP (see below).

\subsubsection{Coordination numbers}

\begin{figure}
\epsfxsize=7cm
\epsfbox{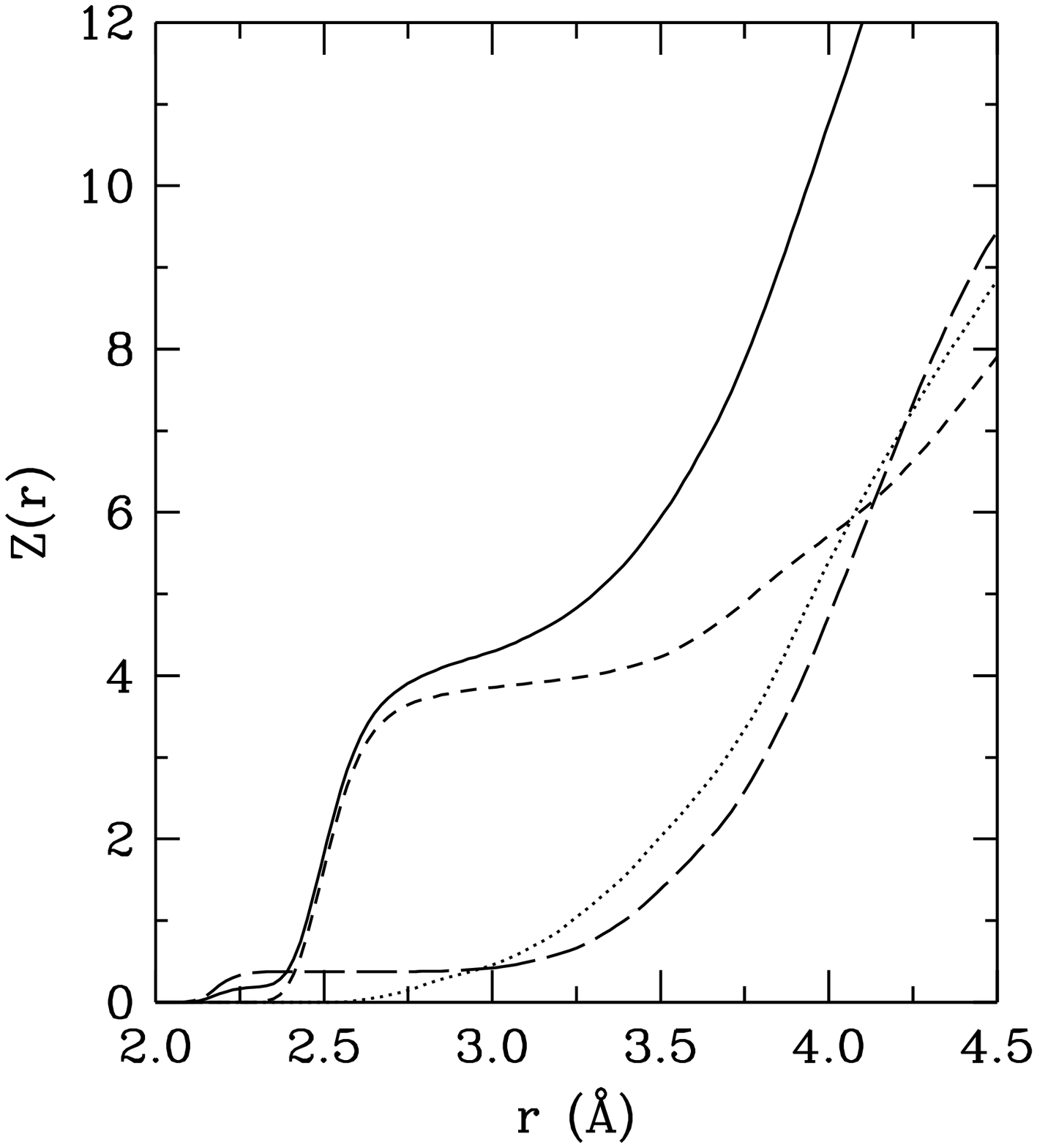}
\caption{
Running coordination numbers corresponding to the radial distribution
functions of Fig.\ \protect\ref{a_rdf}.
\label{a_Zr}
}
\end{figure}

The average coordination numbers can be obtained by integrating the
appropriate RDF's up to the nearest-neighbour distances defined above; this
and other relevant numbers are listed in Table \ref{partc}, while the
``running'' coordination numbers --- now exhibiting plateaux --- are
presented in Fig.\ \ref{a_Zr}. We obtain in this way a total coordination
number of $Z=$ 4.27, in reasonable agreement with the experimental value
mentioned above, $Z \approx 4$, i.e., within the uncertainties inherent to
both methods.

A detailed picture of the short-range structure is provided by the partial
coordination numbers $Z_i$ and $Z_{ij}$, $i,j=$ In or P, also listed in Table
\ref{partc}. We find the partial coordination numbers of In and P to be
almost identical --- 4.25 and 4.29, respectively. {\em Modulo} the
limitations mentioned above, this is again in agreement with the available
experimental values ($4.0 < Z_{\mbox{\scriptsize In-In}} < 4.8$ and
$Z_{\mbox{\scriptsize P-P}} \approx 4.0$). Thus, despite the large
difference in size, and because of the strongly ionic character of the
material, each atom is surrounded by the same average number of atoms. In
this sense, it can be said that all atoms occupy the same volume.

If we detail further the average coordination numbers, we find, from Table
\ref{partc}, that coordination essentially consists of hetero bonding, i.e.,
the system is chemically ordered. Thus, in the case of In, out of the 4.25
neighbours, 3.91 are P and only 0.34 are In. Likewise, for P, which has 4.29
neighbours, we have 3.91 In and 0.38 P. We also see, upon comparing with the
TB-MD results for {\em a-}GaAs, that the chemical short-range order appears
to be a bit stronger in the latter. However, it must be said that the data
reported in Table \ref{partc} (and following) were obtained using the ART
procedure, which allows more extensive relaxation of the network than is
possible with MD.

{\em Modulo} the error bars, the average coordination number is larger in the
amorphous phase than in the ZB crystal at equal density, i.e., there are a
number of overcoordinated atoms. This can be seen in Table \ref{distrib},
where we present the distributions of coordination numbers in our amorphous
sample. Even though the distribution is rather sharply peaked, there are
nevertheless a significant number of coordination defects. In fact, we find,
overall, very few (1.9\%) atoms that are undercoordinated ($Z<4$), while
quite many (26.8\%) are overcoordinated. This, again, contrasts quite sharply
with {\em a-}GaAs, which is slightly undercoordinated; this might be the
cause, in part, for the observation of a lower density in {\em a-}GaAs than
in {\em c-}GaAs.\cite{laaziri94} (Disorder itself is expected to cause a
decrease in density). In contrast, the predominance of overcoordinated
defects in {\em a-}InP is likely responsible for its larger density compared
to {\em c-}InP,\cite{cliche94} given, as we have seen above, that the average
bond length in the amorphous phase is {\em larger} than in the crystal.

\subsubsection{Chemical disorder and wrong bonds}

A quantitative measure of chemical correlations in the binary compound $AB$
is provided by the ``concentration-concentration'' coordination number,
$Z_{cc} = c_B (Z_{AA} - Z_{BA}) + c_A (Z_{BB} - Z_{AB})$ (see, e.g., Ref.\
\onlinecite{elliott83}) where $c_i$ is the concentration of $i$-type atoms in
the system. $Z_{cc}=-4$ exactly in {\em c-}InP; for our amorphous sample, we
find $Z_{cc}=-3.55$ (Table \ref{partc}), indicating, as was already evident
from the above discussion, a rather strong chemical order. Chemical order can
also be quantified in terms of the generalized Warren chemical short-range
order parameter,\cite{elliott83} $\alpha_W = Z_{cc}/(c_B Z_A + c_A Z_B)$,
where $Z_i = \sum_j Z_{ij}$. $\alpha_W=0$ indicates complete randomness
whereas positive and negative values indicate preference for homo and hetero
nearest-neighbour coordination respectively. Evidently, in {\em c-}InP,
$\alpha_W=-1$; for {\em a-}InP, we obtain $\alpha_W=-0.84$ (cf.\ Table
\ref{partc}), revealing, again, a strong preference for chemical ordering, a
bit weaker, perhaps, than in {\em a-}GaAs, for which the TB-MD model gives
$Z_{cc}=-3.54$ and $\alpha_W=-0.88$. This, again, reveals the importance of
Coulombic ordering in InP and GaAs.

The overall similarity between the RDF's of group-IV materials and the III-V
semiconductors suggests that the materials have comparable short-range
structure.\cite{shevchik73} However, as discussed above, there exists a
significant number of coordination defects, such that the overall
coordination exceeds, in the present case, the canonical value of 4.
Likewise, the structure exhibits a significant number of ``anomalous'' rings
--- as can be seen from Table \ref{ring} --- and in particular odd-membered,
just as they can be found in {\em a-}Si or {\em a-}Ge. An immediate
consequence of this is that there must exist ``wrong'' bonds in the
structure. We find in our model that 8.4\% of the bonds are wrong (cf.\ Table
\ref{partc}). Such a proportion of wrong bonds is remarkably small in view of
the fact that the system is slightly overcoordinated and thus is manifest of
the excellent quality of the model.

Experimentally, the proportion of wrong bonds has been reported to lie in the
range 10--40\%.\cite{flank87} The large spread in the values reported is
explained by the fact that some samples are believed to phase separate. It
has been conjectured, also, that the wrong bonds in {\em a-}InP might
originate from local composition fluctuations rather than coordination
defects. Our calculations indicate that coordination defects are responsible
for the wrong bonds. It is perhaps appropriate to remark that it is quite
difficult to imagine an amorphous network {\em without} coordination defects
and/or odd-membered rings, but of course the density of such defects is not
known precisely and probably depends quite strongly on the ``method of
preparation'', be it experimental or computational. In fact, the ``effort''
required to reduce the proportion of wrong bonds to a value smaller than the
present 8.4\% would appear to be formidable if it is a consequence of model
limitations, and in particular the quench rate used in the MD simulations. In
any event, the proportion of wrong bonds we obtain here must be taken as an
upper limit to the actual value: it is certainly the case that the number
would decrease if corresponding simulations were carried out on a larger
system, so as to minimize the elastic constraints, on longer timescales, in
order to allow more complete relaxation. An ART optimization could resolve
the issue;\cite{mousseau97,barkema96} however, this is presently not feasible
{\em ab initio}, or otherwise, since there exists no model potentials for
InP.

\subsubsection{Vibrational properties}

\begin{figure}
\vspace{1cm}
\epsfxsize=7cm
\epsfbox{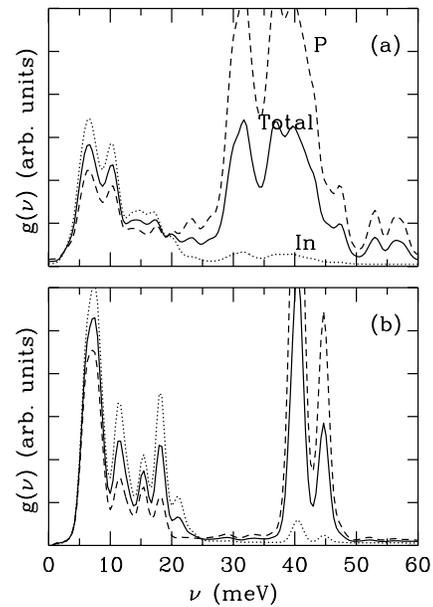}
\vspace{2cm}
\caption{
Densities of vibrational states of (a) {\em a-}InP and (b) {\em c-}InP at 300 K.
\label{a_vdos}
}
\end{figure}

The partial and total densities of vibrational states (DOS) as deduced from
our model are presented in Fig.\ \ref{a_vdos}(a); for reference, we give, in
Fig.\ \ref{a_vdos}(b), the corresponding DOS for {\em c-}InP calculated
within the same computational framework. To our knowledge, there exists no
experimental measurements of this quantity for {\em a-}InP, while the
vibrational spectrum of the crystalline material is well
characterized.\cite{adachi92,landolt1} Thus, in {\em c-}InP, rather wide TA
and LA bands are found in the range 6--9 and 20--23 meV, respectively, while
more sharply defined, TO and LO peaks are seen at about 37 and 42 meV,
respectively. In the present calculation, we find a large TA peak at about
7-8 meV, and a fairly broad LA band in the range 16-22 meV. The sharp optic
peaks are found at 41 and 45 meV, respectively. Thus it appears that our
model overestimates a little bit the energy of the optic peaks, while it
underestimates a little the energy of the LA band. It must be said, however,
that the low-frequency acoustic modes are the most difficult to probe with
molecular dynamics (explaining, in part, the oscillatory structure at low
energies).

The density of states of our model {\em a-}InP agrees, ``broadly'' speaking,
with that of {\em c-}InP, except for a significant softening of the
higher-energy peaks. The total DOS exhibits a well-defined peak at about
7--10 meV, corresponding to the crystal's TA peak, a broad band centered at
about 18 meV, close to the crystal's LA peak, and two well-defined peaks at
32 and 38 meV, corresponding to the crystal's TO and LO bands.

Fig.\ \ref{a_vdos} reveals yet another feature in the DOS which is absent in
the crystal, as well as in elemental semiconductors, namely a (rather broad)
peak at high frequency --- about 55 meV. It is clear from Fig.\ \ref{a_vdos}
that the optic peaks are primarily associated with the fast and energetic
motion of the lightest atom, P, against the heavier one, In. (Cf.\ also the
discussion concerning Fig.\ \ref{l_vdos} in Sec.\ \ref{liquid}.) In view of
this, and of the fact that wrong bonds do not exist in elemental
semiconductors (while other defects, e.g., coordination, do), we conjecture
that the band at 55 meV arises from the motion of phosphorus atoms against
one another, i.e., P-P wrong bonds. (Because of the heavier mass of indium,
In-In wrong bonds will show up at much smaller energies, and therefore be
buried in the continuum of states. For similar reasons, a high-energy
wrong-bond peak has not been observed in the DOS of {\em
a-}GaAs.)\cite{mousseau97} It would be of utmost interest that experimental
confirmation of this point be carried out, since this would give a direct
indication of the presence of wrong bonds and a measure of their relative
importance.

\subsubsection{Electronic properties}

\begin{figure}
\vspace{1cm}
\epsfxsize=7cm
\epsfbox{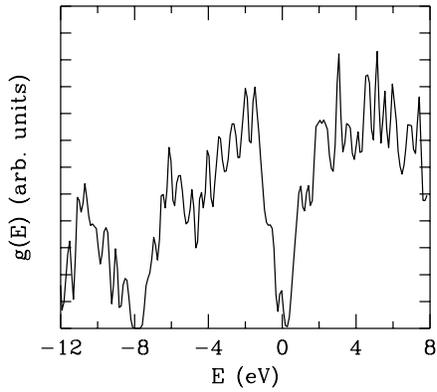}
\caption{
Density of electron states for the amorphous sample at 0 K.
\label{a_edos}
}
\end{figure}

We have calculated the density of electron states $g(E)$ for our model {\em
a-}InP at 0 K; the results are shown in Fig.\ \ref{a_edos}. (In order to
improve presentation --- in view of the limited statistics of the model ---
the density of states has been smoothed lightly using a Gaussian filter of
width 0.15 eV.) The forbidden energy gap, about 1.08 eV, is clearly visible
about the Fermi energy. For crystalline InP in the ZB phase, our computed
value for the LDA direct gap at the $\Gamma$ point is 1.50 eV (at $a$=
5.68~\AA) and compares well with the value 1.50~eV of Ref.\
\onlinecite{dalcorso} and with experiment, 1.42 eV. The gap of {\em a-}InP,
therefore, is a bit smaller than that of the crystalline material.

One important difference, however, is that there are defect states in the gap
of {\em a-}InP not present in (ideal) {\em c-}InP. Thus, we have identified
one particular electron level giving rise to contributions near mid-gap in
$g(E)$, clearly visible in Fig.\ \ref{a_edos}. We have examined the local
density of states for this particular level, and found that it corresponds to
an empty, distorted, octahedral, mostly-indium (five out of six corners),
``cage'', i.e., basically, a cluster of wrong In-In bonds. That wrong bonds
give rise to states in the gap has also been inferred from a comparison of a
Polk-type (with wrong bonds) with a Connell-Temkin-type (without wrong bonds)
model for {\em a-}GaAs, as discussed above (Sec.\ \ref{backT}).

\section{Can amorphization be pressure-induced?}\label{pressure}

In an attempt to verify the possibility that InP could amorphize under
compression, we subjected the equilibrium ZB structure to pressure by
increasing ``slowly'', in steps, the density. Referring to Fig.\
\ref{E_P_vs_a}(b), we find the correspondence between pressure and density
(in fact the lattice parameter). Starting with InP in its perfect ZB
arrangement, properly equilibrated at 300 K, the lattice parameter was thus
decreased (always at 300 K) from 5.68 \AA\ (equilibrium) to 5.23 \AA, i.e.,
down to a value smaller than (cf. Sec.\ III) the equilibrium lattice constant
for the NaCl phase. From Fig.\ref{E_P_vs_a}(a), we would expect the ZB
crystal to undergo a transition to the high-pressure NaCl phase in the MD run
when $a$ is set to values below about 5.4 \AA. (This value corresponds to a
pressure of about 140 kbar; at the highest density investigated here, for
$a=5.23$ \AA, the pressure in the ZB structure is 270 kbars [cf.\ Fig.\
\ref{E_P_vs_a}(b)].)

\begin{figure}
\vspace{1cm}
\epsfxsize=7cm
\epsfbox{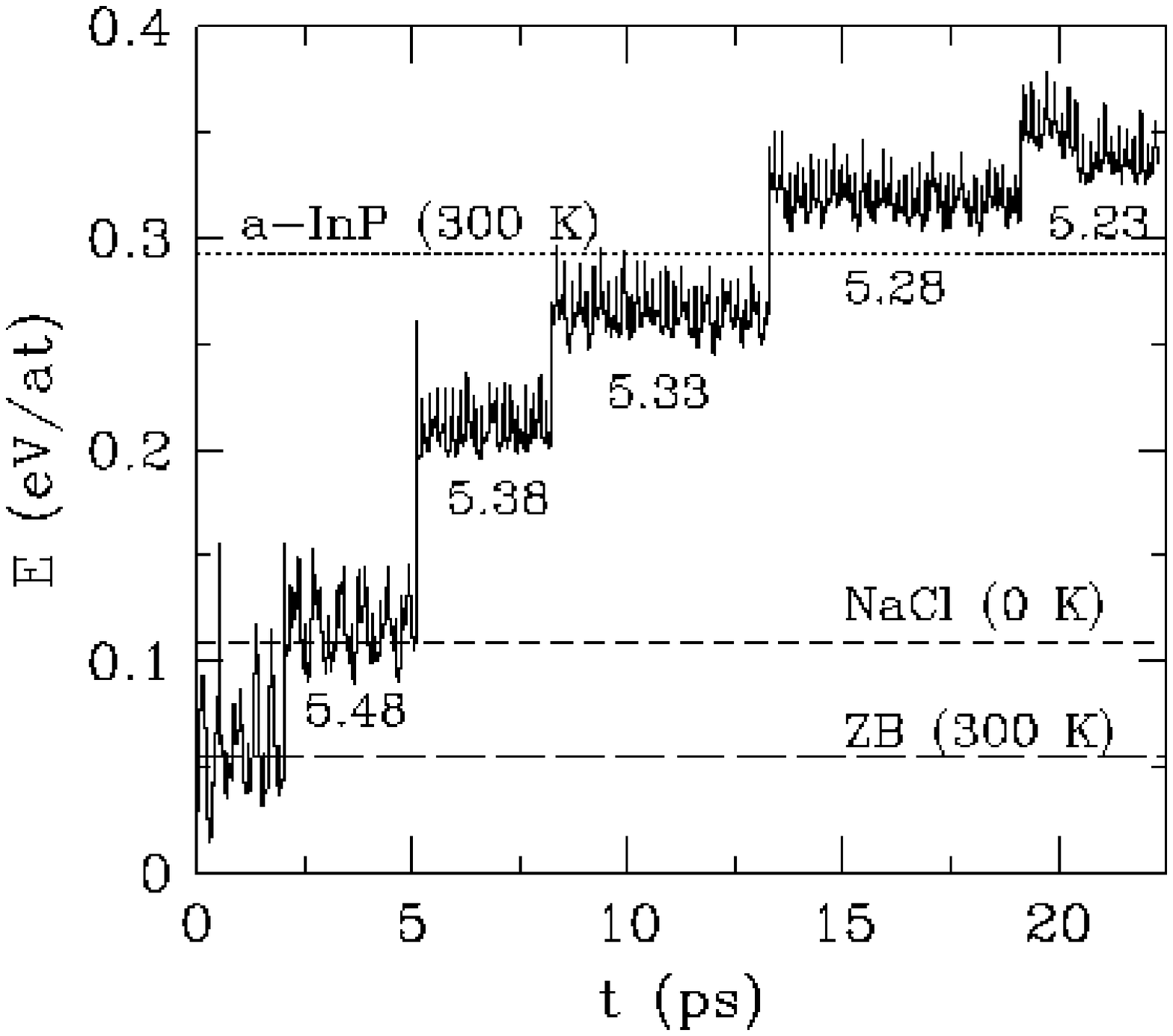}
\caption{
Energy versus time-density for the compressed ZB phase at 300 K relative to
the ZB crystal at 0 K. Also indicated are the energies of the NaCl phase at 0
K, of the equilibrium ZB phase at 300 K, and of the amorphous phase at the
equilibrium density.
\label{p_etot}
}
\end{figure}

\begin{figure}
\epsfxsize=10cm
\epsfbox{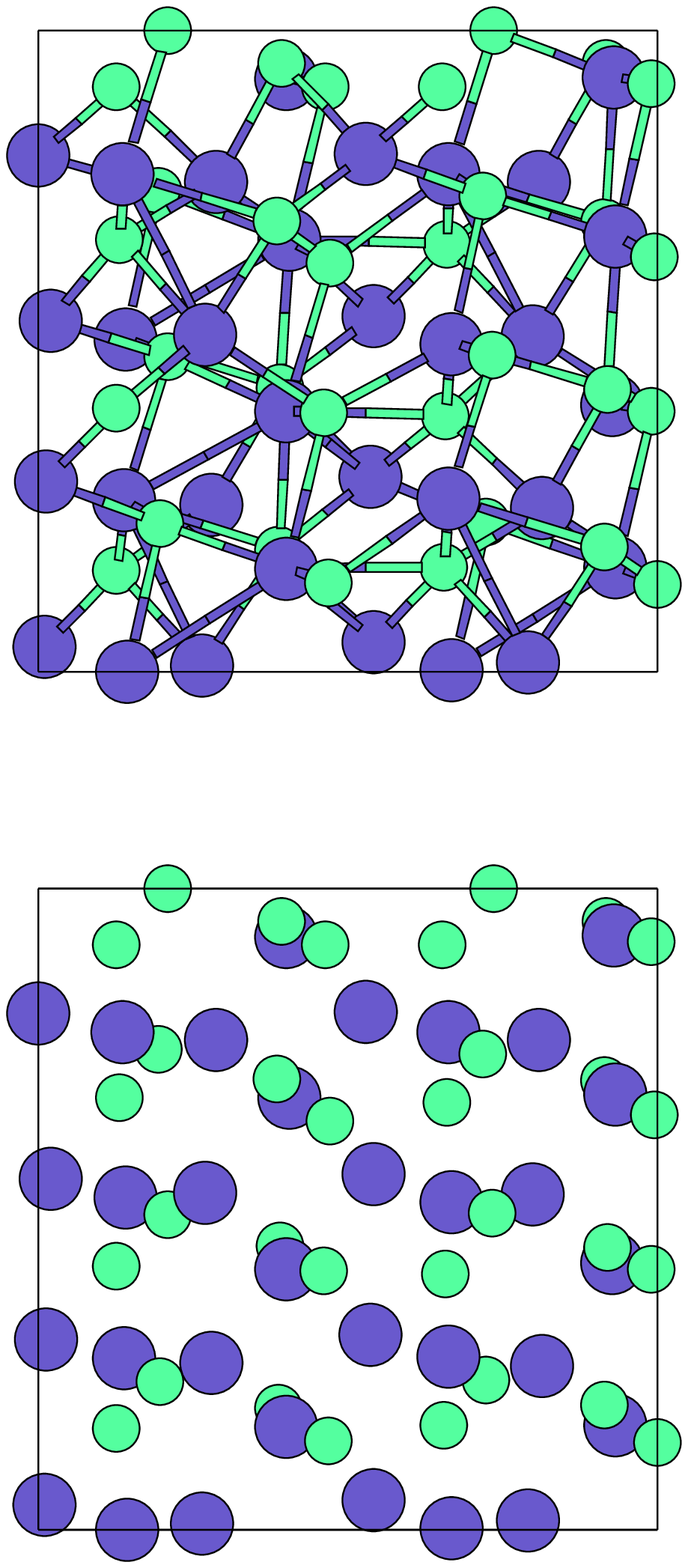}
\caption{
Top: Ball-and-stick representation of the final, compressed ($a=5.23$ \AA),
ZB crystal; bottom: same, with bonds removed in order to show better the
underlying structure.
\label{p_xmol}
}
\end{figure}

Fig.\ \ref{p_etot} shows the evolution of the total energy of the system at
300 K relative to the ZB crystal at 0 K, as a function of time-density. For
reference, we also show on this plot the total energy of the equilibrium ZB
phase at 300 K, the total energy of the amorphous phase (also at 300 K) {\em
at the equilibrium density}, and the total energy of the NaCl phase at 0 K,
obtained as discussed earlier. We find that the energy increases rather
smoothly with density. At a value of $a=5.33$ \AA, visual inspection of the
system indicates that it undergoes some sort of distortion into a state which
is definitely not ZB, but which bears strong resemblence to it. This
distortion is also visible in Fig.\ \ref{p_etot} as a slight decrease of the
total energy as a function of time. At density values corresponding to
lattice parameters smaller than about 5.28 \AA, the energy of the compressed
ZB crystal exceeds that of the amorphous phase. Yet, no transition to an
amorphous phase takes place. Upon increasing the density further, we observe
another transformation for $a=5.23$ \AA, clearly visible in Fig.\
\ref{p_etot} --- the energy drops significantly, to a state which is
evidently distinct from the NaCl structure (its computed energy is much
higher). We have not analyzed this phase in detail but it is evidently
ordered, as can be seen in Fig.\ \ref{p_xmol}, and might possibly be an
intermediate state on the way to the NaCl phase. One thing is clear, however:
the new phase maintains the chemical order of the system, i.e., introduces no
wrong bonds. It would seem, therefore, that pressure is {\em not} a proper
route for amorphization; rather, the system prefers to reorganize into a new
crystalline form, which is more favorable in view of the high cost in energy
of wrong bonds.

The argument might be presented in another way: In Fig.\ \ref{p_etot}, we see
that the energy of the new phase, 0.33 eV/at, lies only slightly above that
of the amorphous phase obtained by the melt-and-quench cycle, 0.29 eV/at. The
latter value, however, is at the {\em equilibrium} density. Under
compression, the energy of the amorphous phase would also go up, presumably
by an energy smaller than but comparable to 0.33~eV/at, and thus would
largely exceed that of the compressed crystal. In view of this, it is very
difficult to imagine that amorphization could take place under compression.
Rather, a transition to the NaCl phase would take place.

The emerging picture suggested by our simulations is that InP does {\em not}
amorphize under pressure, even at values largely exceeding those required for
the system to transform into the high-pressure NaCl phase. The energy of the
amorphous phase lies well above that of the compressed crystal and the cost
of breaking the strong ionic bonds is just too large. The system, rather,
finds its way into a complex, but {\em ordered}, phase that maintains the
chemical order of the system, i.e., that does not introduce wrong bonds. It
appears, therefore, that, indeed, strongly-ionic materials are not good
candidates to pressure-induced amorphization:\cite{tsuji95} more ``violent''
processes (such as implantation) are required, and this suggests that the
amorphous phase cannot exist in absence of wrong bonds. Though we have not
examined this, it is not impossible that InP would amorphize from the
high-pressure phase upon the release of pressure and/or through proper
heat/pressure treatment. In view of the above energy-wise arguments, however,
this seems to be very unlikely.

\section{Concluding Remarks}\label{concl}

We have carried a detailed and extensive first-principles molecular-dynamics
study of the structure and electronic properties of amorphous InP produced by
rapid quenching from the liquid. The structure of the material is found to be
strongly ordered chemically, about the same, for instance, as in {\em
a-}GaAs, even though there are a significant number of coordination defects
(antisites) and despite the presence of odd-membered rings. We find, as a
consequence, that there exists ``wrong bonds'' in the system, in an amount of
about 8\%; these are a consequence of the presence of defects in the system,
{\em not} of composition fluctuations, as has been conjectured. The system,
in fact, is found to be over-coordinated, which might be the reason for the
observed higher density of {\em a-}InP compared to {\em c-}InP. We have also
investigated the possibility of pressure-amorphizing InP. Our calculations
indicate that the cost of a transformation of the compressed zinc-blende
crystal into an amorphous phase is so large that it is very unlikely that it
would take place.

\acknowledgements

It is a pleasure to thank Dr.\ Normand Mousseau and Prof.\ S. Roorda, for
useful comments and critical reading of this manuscript. LJL is grateful to
Prof.\ Car for hospitality and support at IRRMA where most of the work was
carried out. Support from the Natural Sciences and Engineering Research
Council (NSERC) of Canada and the ``Fonds pour la formation de chercheurs et
l'aide \`a la recherche'' of the Province of Qu\'ebec is also gratefully
acknowledged.


\newpage

\begin{center}
\begin{table}
\caption{
Experimental values of the nearest-neighbour distances and partial
coordination numbers. For In$_{33}$P$_{67}$, the results of two different
fits to the same EXAFS data are indicated.
}
\label{exp_st}
\begin{tabular}{lccccccccc}
Sample & \multicolumn{4}{c} {$r_{NN}$ (\AA)} & \multicolumn{4}{c} {$Z$} & Ref.
\\ \hline
                 & In-In & In-P & P-In & P-P  & In-In & In-P & P-In & P-P & \\
\hline
{\em a-}In$_{33}$P$_{67}$ & 2.98  & 2.59 & 2.58 & 2.20 &  1.5  & 2.5  & 1.8  & 2.2 &
\protect\onlinecite{flank87} \\
                  &       &      &      &      &  1.8  & 3.0  &      &     &
\protect\onlinecite{flank87} \\
{\em a-}In$_{35}$P$_{65}$ & 2.80  & 2.58 & 2.58 & 2.24 &  1.2  & 2.8  & 1.6  & 2.5 &
\protect\onlinecite{udron92} \\
{\em a-}In$_{40}$P$_{60}$ & 2.76  & 2.57 & 2.58 & 2.24 &  0.9  & 3.1  & 2.2  & 1.8 &
\protect\onlinecite{udron92} \\
\end{tabular}
\end{table}
\end{center}

\begin{center}
\begin{table}
\caption{
Diffusion constants in the liquid at various temperatures, in units of
10$^{-4}$ cm$^2$/s.
}
\label{diff}
\begin{tabular}{lc}
$T$ (K) & $D$  \\ \hline
3000    & 2.13 \\
2700    & 2.32 \\
2400    & 1.96 \\
2100    & 1.43 \\
\end{tabular}
\end{table}
\end{center}

\begin{center}
\begin{table}
\caption{
Partial and total coordination numbers in the liquid; the cutoff distances
$r_Z$ are also given. The total $Z$ is obtained from the partials as $Z =
\sum_{ij} c_i Z_{ij}$.
}
\label{tab_Z}
\begin{tabular}{lccccccc}
$T$ (K) & \multicolumn{2}{c} {In-In} & \multicolumn{2}{c} {In-P} &
\multicolumn{2}{c} {P-P} & Total \\
        & $r_Z$ & $Z$ & $r_Z$ & $Z$ & $r_Z$ & $Z$ & $Z$ \\ \hline
3000    & 4.0   & 5.6 & 3.7   & 5.1 & 2.8   & 1.6 & 8.7 \\
2700    & 4.0   & 5.6 & 3.6   & 4.8 & 2.8   & 1.8 & 8.5 \\
2400    & 3.7   & 4.2 & 3.6   & 4.8 & 2.7   & 1.5 & 7.7 \\
2100    & 3.5   & 3.4 & 3.5   & 4.6 & 2.7   & 1.4 & 7.0 \\
\end{tabular}
\end{table}
\end{center}

\begin{center}
\begin{table}
\caption{
Number per atom of $n$-membered rings for the {\em a-}InP sample at 300 K, as
well as for the ideal {\em c-}InP structure. Also shown, for comparison, are
the results for {\em a-}GaAs obtained from a fully-relaxed, ART-optimized,
TB-MD model.\protect\cite{mousseau97}
}
\label{ring}
\begin{tabular}{lccccc}
 $n$                    &  3    &  4   &  5   &  6   &  7   \\ \hline
    {\em c-}InP         &  0    &  0   &  0   &  4   &  0   \\
    {\em a-}InP         &  0.02 & 0.44 & 0.37 & 2.35 & 4.37 \\
    {\em a-}GaAs        &  0.05 & 0.10 & 0.21 & 1.37 & 0.76 \\
\end{tabular}
\end{table}
\end{center}

\begin{center}
\begin{table}
\caption{
Structural properties of {\em a-}InP at 300 K: coordination numbers $Z$
(partial, species, total, concentration-concentration), Warren chemical
short-range order parameter $\alpha_W$, and proportion of wrong bonds (WB).
Also shown, for comparison, are the results for {\em a-}GaAs obtained from a
fully-relaxed, ART-optimized, TB-MD model.\protect\cite{mousseau97} Here $A$
represents either In or Ga and $B$ represents either P or As.
}
\label{partc}
\begin{tabular}{lccccccccc}
             & $Z_{AA}$ & $Z_{AB}$ & $Z_{BB}$ & $Z_{A}$ & $Z_{B}$ & $Z$  & $Z_{cc}$ & $\alpha_W$ & WB   \\ \hline
{\em c-}InP  &   0       & 4         & 0         & 4       & 4       & 4    & $-$4.00  & $-$1.0     & 0    \\
{\em a-}InP  &  0.34     & 3.91      & 0.38      & 4.25    & 4.29    & 4.27 & $-$3.55  & $-$0.84    & 8.4  \\
{\em a-}GaAs &  0.22     & 3.75      & 0.21      & 3.97    & 3.96    & 3.96 & $-$3.54  & $-$0.88    & 5.2  \\
\end{tabular}
\end{table}
\end{center}

\begin{center}
\begin{table}
\caption{
Distribution (in \%) of total coordination numbers for the {\em a-}InP sample
at 300 K, as well as for the ideal {\em c-}InP structure. Also shown, for
comparison, are the results for {\em a-}GaAs obtained from a fully-relaxed,
ART-optimized, TB-MD model.\protect\cite{mousseau97}
}
\label{distrib}
\begin{tabular}{lcccccccc}
$Z$           & 0 & 1 & 2 & 3    &  4   &  5   & 6   & 7   \\ \hline
{\em c-}InP   & 0 & 0 & 0 & 0    & 100  &  0   & 0   & 0   \\
{\em a-}InP   & 0 & 0 & 0 & 1.9  & 71.3 & 24.7 & 2.1 & 0   \\
{\em a-}GaAs  & 0 & 0 & 0 & 11.1 & 82.8 &  5.2 & 0.6 & 0.2 \\
\end{tabular}
\end{table}
\end{center}

\newpage

\begin{figure}
\epsfxsize=7cm
\epsfbox{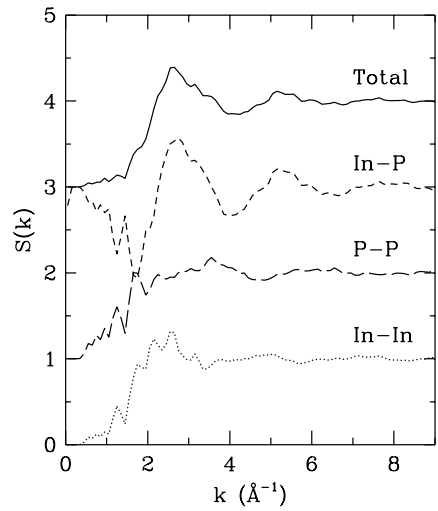}
\caption{
Partial and total static structure factors of liquid InP at 2100 K.
\label{l_ssf}
}
\end{figure}

\end{document}